\documentclass[twocolumn,superscriptaddress,showpacs,pra]{revtex4}
\usepackage{mathrsfs}
\usepackage{amssymb, amsbsy, amsmath, latexsym, dsfont, array, layout, graphicx,mathrsfs,color}
\usepackage{graphicx}
\usepackage{dcolumn}
\usepackage{bm}

\newcommand{\ket}[1]{\left|{#1}\right\rangle}
\newcommand{\bra}[1]{\left\langle{#1}\right|}

\begin{document}

\preprint{APS/123-QED}

\title{Linear optical demonstration of quantum speed-up with a single qudit}

\author{Xiang Zhan}
 \address{Department of Physics, Southeast University, Nanjing, 211189, China.}
\author{Jian Li}
\address{Department of Physics, Southeast University, Nanjing, 211189, China.}
\author{Hao Qin}
\address{Department of Physics, Southeast University, Nanjing, 211189, China.}
\author{Zhihao Bian}
\address{Department of Physics, Southeast University, Nanjing, 211189, China.}
\author{Peng Xue}
\email{gnep.eux@gamil.com}
\address{Department of Physics, Southeast University, Nanjing, 211189, China.}
\affiliation{State Key Laboratory of Precision
Spectroscopy, East China Normal University, Shanghai 200062, China}
\date{\today}

\begin{abstract}
Though quantum algorithm acts as an important role in quantum computation science, not only for providing a great vision for solving classically unsolvable problems, but also due to the fact that it gives a potential way of understanding quantum physics, the origin of the power of quantum algorithm is still an open question. Non-classical correlation is regarded as the most possible answer for the open question. However we experimentally realize a quantum speed-up algorithm on four-level system with
linear optical elements and prove that even a single qudit is enough for designing an
oracle-based algorithm which can solve a
model problem twice faster than any classical algorithm. The algorithm can be generalized to higher dimensional qudits with the
same two to one speed-up ratio.
\end{abstract}

\maketitle


{\it Introduction}: - Harnessing the miraculous features of quantum physics, quantum computation based on well-designed quantum algorithms shows great advantages over their classical counterparts. Due to the magic power of quantum algorithms, e.g. Shor's algorithm to factor large numbers \cite{Shor,Lieven}, quantum simulation algorithm \cite{Feynman,Lloyd,Aaronson,Broome,Xiaoqi}, a fascinating algorithm which is realized recently to solve Simon's problem \cite{simontheo,simontheo2,Simon,Long}, and the more recent algorithm used to solve linear equations \cite{Harrow,Cai,Du}, quantum algorithms have drawn much attention of researchers. Development of quantum algorithms has also proposed an unsolved open question: what actually is the resource that acts as the intrinsic origin of the vigoroso power of quantum algorithms \cite{Nest,Howard,Lanyon}. For decades, there are many possible answers for the question: supposition, entanglement, quantum discord, and contextuality \cite{Howard,Klyachko}. Specific algorithms which make use of specific properties are promising for revealing the relation between these properties and quantum speed-up.

Recently, Gedik \cite{Gedik} proposed an interesting and simple quantum algorithm which gives an intuitive computational speed-up without requirements of any kind of quantum correlations. The designed algorithm intends to determine the parity of a given permutation function of a set by performing the permutation operation only once, which shows a two-to-one speed-up ratio over the corresponding classical algorithm. The algorithm has been reported to be realized in nuclear magnetic resonance system \cite{NMR1,NMR2}. In this paper, we report an experimental demonstration of the algorithm for four-dimensional set with linear optical system. The algorithm can be generalized to higher dimensional cases with the
same speed-up ratio.

{\it Gedik's algorithm}: - Gedik's algorithm solves a black-box problem as follows: considering a set S=\{0,1,...,d-1\} with {\it d} elements (a {\it d}-dimensional set), the black-box is one of the permutations defined on the set, which maps these {\it d} inputs into {\it d} possible outputs. The permutation function can be summarized as functions of the elements,
 \begin{equation}
 f_{m}^{\pm}(x)=(m\pm x)_{\text{mod}(d)},
 \label{eq:1}
 \end{equation}
 where $x\in S$, $m=0,1,..d-1$, and mod({\it d}) denotes that $(m\pm x)$ is taken modulo {\it d}. For a given permutation, there is a global property named parity. The parity is positive (or negative) if the permutation function has the form $f_{m}^{+}(x)$ (or $f_{m}^{-}(x)$). The task is to determine the parity of a permutation, which is a global property of the black-box. Classically, one needs to evaluate the permutation function at least twice for different inputs to get the parity. Whereas with Gedik's algorithm, we show below that performing the function only once is enough.

\begin{figure}[htb]
  \includegraphics[width=8cm]{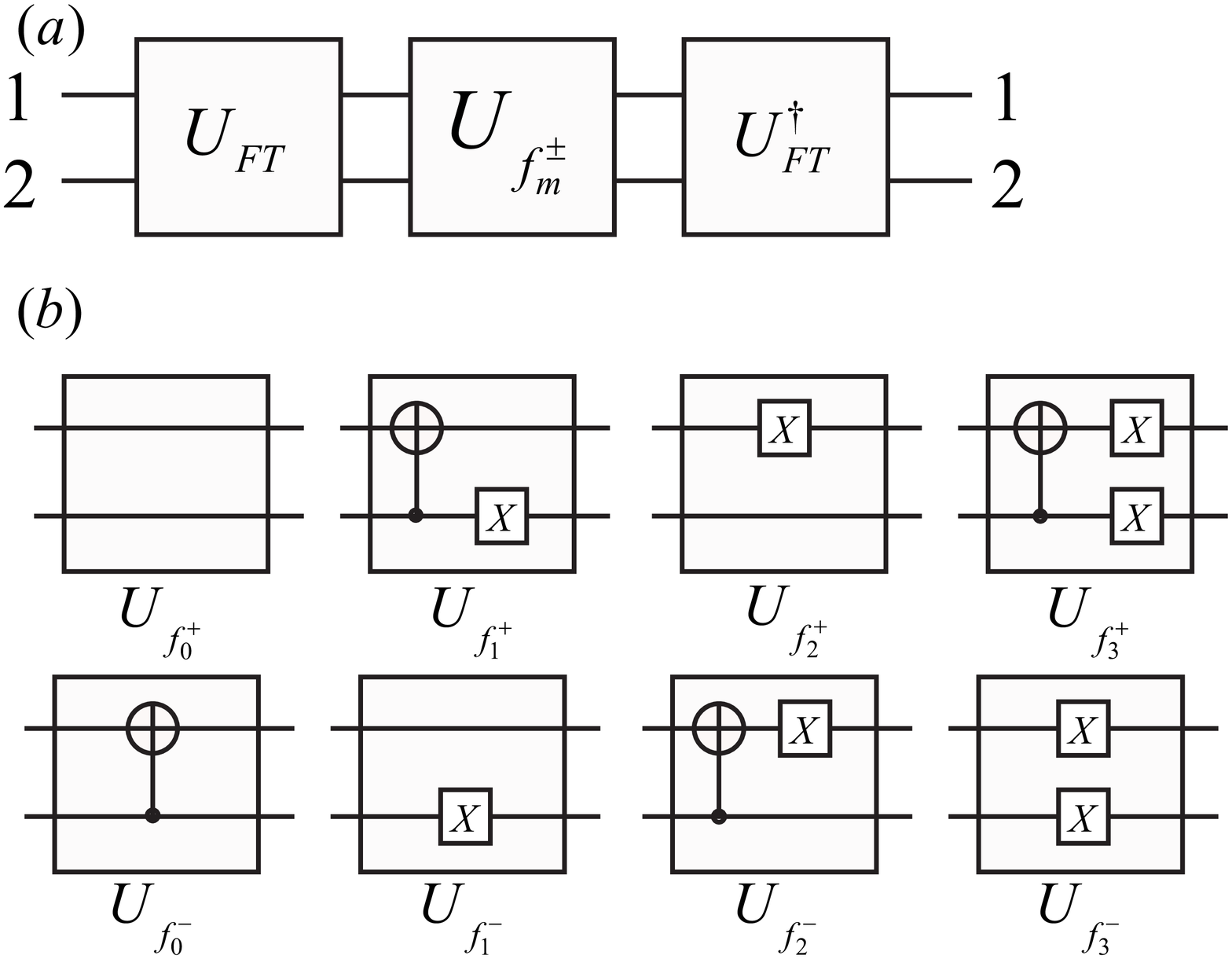}
  \caption{The circuit model of Gedik's algorithm: - The upper line denotes the first qubit, and the lower one for second. (a) The global circuit of the algorithm for d=4 case. The squares represent three modules of the algorithm (see main text for details). (b) Circuit for the permutation function of a qudit state. It is easy to see that all the operations can be constructed by CNOT and {\it X} gates. The circuit for $U_{f_{3}^{+}}$ is not the simplest choice, but we design it to avoid introducing any extra single-qubit gate in front of CNOT gate, which will cause the photons into the CNOT module to be time distinguishable. }
  \label{fig:1}
  \end{figure}

In order to solve the problem with a quantum algorithm, the elements of a {\it d}-dimensional set correspond to a complete orthogonal basis states $\{\ket{0}$,...,$\ket{d-1}\}$ of a {\it d}-dimensional Hilbert space, where $\ket{j}=\{\delta_{(0,j)},\delta_{(1,j)},...,\delta_{(d-1,j)}\}^{T}$. The quantum algorithm contains the following three subroutines (see Fig. \ref{fig:1} (a) for 4-dimensional case): (\romannumeral1) quantum Fourier transform $U_{FT}$, (\romannumeral2) permutation operation $U_{f_{m}^{\pm}}$, (\romannumeral3) inverse quantum Fourier transform $U_{FT}^{\dagger}$.

The initial state of the quantum algorithm is initialized to be $\ket{\psi}_{0}=\ket{1}$. Step (\romannumeral1) is to apply quantum Fourier transform \cite{Nielsen} on it, which results in
\begin{equation}
\ket{\psi}_{1}=\frac{1}{\sqrt{d}}\sum_{d-1}^{k=0}{e^{2\pi ik/d}\ket{k}}.
\label{eq:2}
\end{equation}

\begin{figure*}[htb]
 \includegraphics[width=16cm]{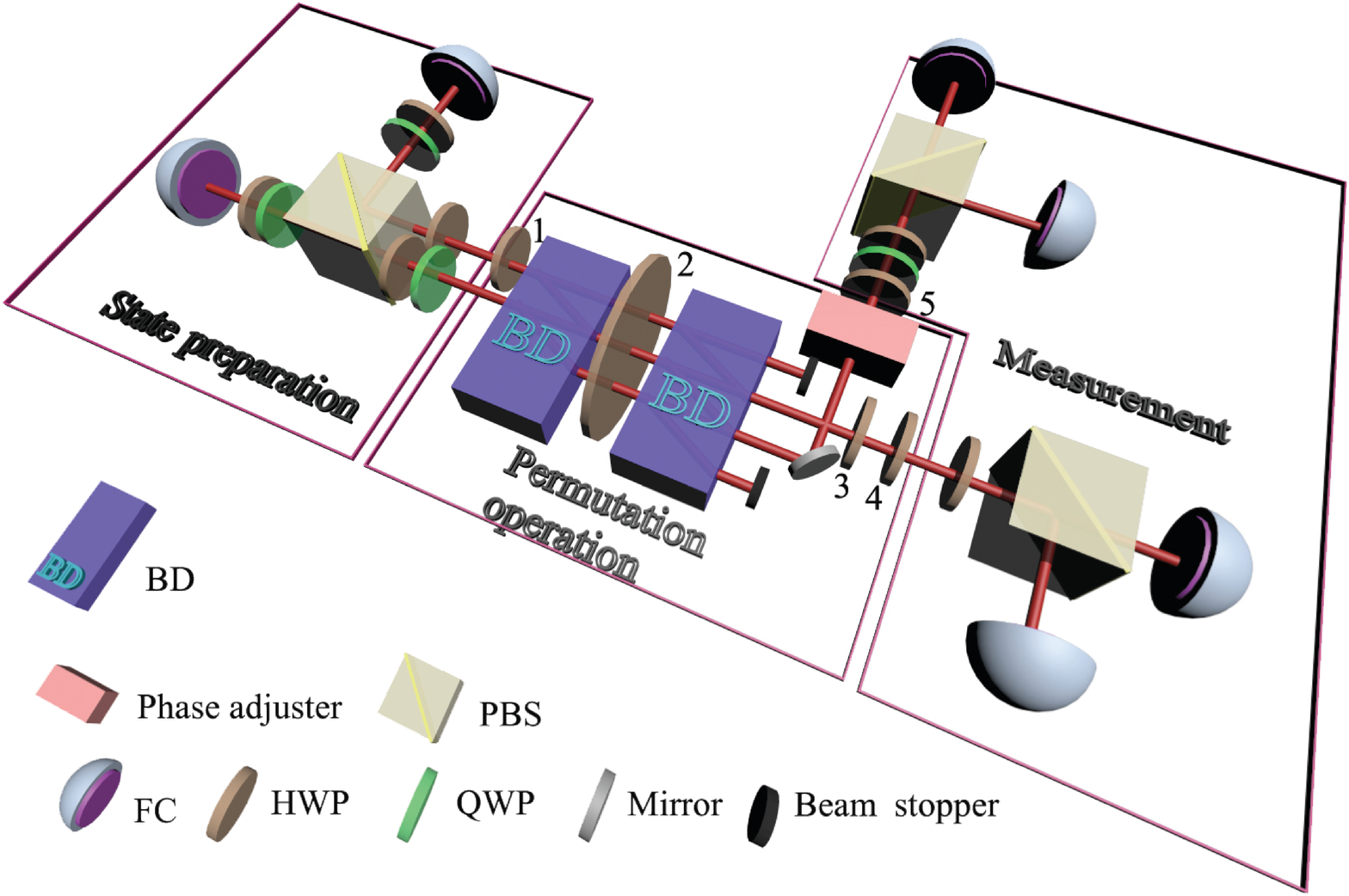}
 \caption{Experimental setup: - The pairs of photons are created via pumping a 0.5 mm-thick nonlinear-$\beta$-barium-borate crystal with a 400.8 nm CW diode laser with 80 mW of power through type-$\uppercase\expandafter{\romannumeral1}$ spontaneous parametric down-conversion, collected by single mode fibers (not shown). There is also a tunable module (not shown) in front of the  optical circuit, which is used to tune the time difference between two photons of one pair. The photons are tuned to be time indistinguishable and then incident into the optical computation circuit via the two fiber collimators (FCs) on the left. The numbers mark the HWP near it for convenient in expression. There are three modules in the setup. (1) State preparation: HWPs and QWPs in front of the first PBS are used to prepare pairs of photons in the state $\ket{HV}$, so that the first photon can be reflected by the PBS, while the second photon is transmitted. Thus pairs of photons propagate parallelly. Two HWPs and a QWP behind the first PBS are for preparing the state of Eq. (\ref{eq:3}). (2) Permutation operation: the CNOT submodule contains two BDs and three HWPs (HWP$_{1}$, HWP$_{2}$ and HWP$_{3}$). HWP$_{1}$ and HWP$_{3}$
  are fixed at 22.5$^\circ$ and -22.5$^\circ$, respectively. When HWP$_{2}$ turns to be 17.5$^\circ$, the CNOT submodule is a CNOT gate with the second qubit the control qubit and the first qubit the target qubit. Whereas, tuning the HWP$_{2}$ to be 45$^\circ$ makes the CNOT submodule operation as {\it X} gates on each qubit. HWP$_{4}$ and  HWP$_{5}$ are removable and fixed at 45$^\circ$. By moving one of the HWPs one can perform an {\it X} gate on the qubit. Tuning the angle of HWP$_{2}$ and moving the HWP$_{4}$ and  HWP$_{5}$ give the eight permutation operations. (3) Measurement: WPs are used to apply semiclassical inverse quantum Fourier transform. The photons are projected into polarization basis via PBS and collected into single mode fibers. The photons are detected by APDs. The result is obtained after a logic coincident count, with coincidence window set at about 1.9 ns.}
\label{fig:2}
\end{figure*}

The permutation operation
\begin{equation}
U_{f_{m}^{\pm}}=\sum_{j=0}^{d-1}{\ket{(m\pm j)_{{\text mod} (d)}}\bra{j}}\notag
 \end{equation}
 is a quantum operator to realize the permutation of Eq. (\ref{eq:1}). In step (\romannumeral2), we apply $U_{f_{m}^{\pm}}$ to the resulting state of step (\romannumeral1), and get the permutated state
\begin{equation}
\ket{\psi}_{2}=\frac{1}{\sqrt{d}}\sum_{d-1}^{k=0}{e^{2\pi ik/d}\ket{(m\pm k)_{{\text mod} (d)}}}.\notag
\end{equation}

The final step is to apply inverse quantum Fourier transform on the system, which is a process to result in the parity of the permutation. Simple calculation shows the final state is dependent on the parity of the permutation. That is, if the parity of the permutation is positive (say the permutation with the form $U_{f_{m}^{+}}$), the algorithm ends up with
\begin{equation}
\ket{\psi}_{3}^{+}=e^{-2\pi im/d}\ket{1},\notag
\end{equation}
otherwise, the negative permutation $U_{f_{m}^{-}}$ makes the algorithm concluding with
\begin{equation}
\ket{\psi}_{3}^{-}=e^{-2\pi i(d-1)m/d}\ket{d-1}.\notag
\end{equation}

Thus, the final state is quite dependent on the parity of the permutation, and the two different outcomes corresponding to two kinds of permutations are orthogonal. Thus performing a projective measurement of the final state is able to determine the parity of the permutation. The parity is positive (or negative) if the outcome of measurement is $\ket{1}$ (or $\ket{d-1}$). It is obvious to see that one can successfully determine the parity of a given permutation by applying the permutation operation only once. This shows a speed-up over the classical algorithm, in which to evaluate the permutation function for two different inputs is necessary. Moreover, the outcome is deterministic, and different outcomes are orthogonal. Thus the projective measurement ensures the probability of successfully getting the parity is unit for ideal quantum circuit.

{\it Experimental realization}: - In this paper, we demonstrate a proof-of-principle experiment of the algorithm for $d=4$ case with linear optical quantum circuit. The four orthogonal states $\ket{j}$ of qudit are represented by binary representation of two-qubit state, i.e., $\ket{0}:=\ket{0}\otimes\ket{0}$, $\ket{1}:=\ket{0}\otimes\ket{1}$, $\ket{2}:=\ket{1}\otimes\ket{0}$, $\ket{3}:=\ket{1}\otimes\ket{1}$, where $\otimes$ denotes tenser product. The first step of the algorithm is to apply quantum Fourier transform to the state $\ket{01}$, which yields
\begin{equation}
(\ket{0}-\ket{1})(\ket{0}+i\ket{1})/2,
\label{eq:3}
\end{equation}
according to Eq. (\ref{eq:2}). It shows that this process maps a product state $\ket{01}$ to another product state in Eq. (\ref{eq:3}). Thus we integrate the quantum Fourier transform into the state preparation module, instead of constructing the original form of quantum Fourier transform circuit, which contains controlled two-qubit gate \cite{Nielsen}. In other words, we prepare the state in Eq. (\ref{eq:3}) directly instead of preparing $\ket{11}$ followed by a quantum Fourier transform. The module of performing permutation operations can be constructed by controlled-NOT (CNOT) gate and bit flipping ({\it X}) gates (see Fig. \ref{fig:1}(b)). Finally, the inverse Fourier transform module is realized in a widely-used semiclassical way \cite{Griffiths,Cai}. Instead of fabricating an inverse quantum Fourier transform circuit with two-qubit controlled gate \cite{Nielsen}, this semiclassical method requires only single-qubit gates performed together with the feedback of classical signals.

In this experiment, the logic qubits $\ket{0}$ and $\ket{1}$ are encoded in the horizontal ($\ket{H}$) and vertical ($\ket{V}$) polarization states of single photons, respectively. Thus the two-photon polarization states form four orthogonal bases. To implement the quantum circuit shown in Fig. \ref{fig:1}, we prepare pairs of separated single photons via pumping a 0.5 mm-thick
nonlinear-$\beta$-barium-borate crystal with a 400.8 nm CW diode laser with 80 mW of power through type-\uppercase\expandafter{\romannumeral1} spontaneous parametric down-conversion.  The experimental challenge mainly lies in the CNOT gate for some permutation operations (see Fig. \ref{fig:1} (b)).

In the state preparation module, which has integrated the quantum Fourier transform, a polarization beam splitter (PBS) and wave plates (WPs) are used to initialize the two-photon state to be Eq. (\ref{eq:3}) (see Fig. \ref{fig:2} for details of experimental setup). The permutation modules are constructed by the circuits shown in Fig. \ref{fig:1} (b). Following the idea of O'Brein et al. \cite{Brien}, the CNOT gate is constructed in an inherently stable architecture by two beam displacers (BDs) and three half wave plates (HWPs) (we prefer to name the architecture formed by those elements ``CNOT submodule''). The {\it X} gate on single qubit is implemented via removable HWPs (HWP$_{4}$ and HWP$_{5}$, where HWP$_{i}$ is the HWP with {\it i} marked in Fig. \ref{fig:2}) set at $45^\circ$. In order to keep the stability of the optical circuit, we do not remove the CNOT submodule when CNOT is not needed for the permutation operations. Instead, we adopt a more stable and efficient method. We turn the angle of HWP$_{2}$ from $17.5^\circ$ for the CNOT gate to $45^\circ$ for two {\it X} gates on both photons. Thus all the eight permutations can be achieved via tuning the angle of HWP$_{2}$, and moving HWP$_{4}$ and HWP$_{5}$ (see Table. \ref{table:1}).

\begin{table}[htbp]
\begin{tabular}{c|c|c|c}
\hline
Permutation operation & $HWP_{2}$ & $HWP_{4}$ & $HWP_{5}$\\
\hline\hline
$U_{f_{0}^{+}}$ & $45^\circ$ & $45^\circ$ & $45^\circ$\\
\hline
$U_{f_{1}^{+}}$ & $17.5^\circ$ & $\setminus$ & $45^\circ$\\
\hline
$U_{f_{2}^{+}}$ & $45^\circ$ & $\setminus$ & $45^\circ$\\
\hline
$U_{f_{3}^{+}}$ & $17.5^\circ$ & $45^\circ$ & $45^\circ$\\
\hline
$U_{f_{0}^{-}}$ & $17.5^\circ$ & $\setminus$ & $\setminus$\\
\hline
$U_{f_{1}^{-}}$ & $45^\circ$ & $45^\circ$ & $\setminus$\\
\hline
$U_{f_{2}^{-}}$ & $17.5^\circ$ & $45^\circ$ & $\setminus$\\
\hline
$U_{f_{3}^{-}}$ & $45^\circ$ & $\setminus$ & $\setminus$\\
\hline
\end{tabular}
\caption{Angles of HWPs to realize different permutation operations, where ``$\setminus$" denotes that the HWP is removed from the optical circuit.}
\label{table:1}
\end{table}

\begin{figure*}[htb]
 \includegraphics[width=16cm]{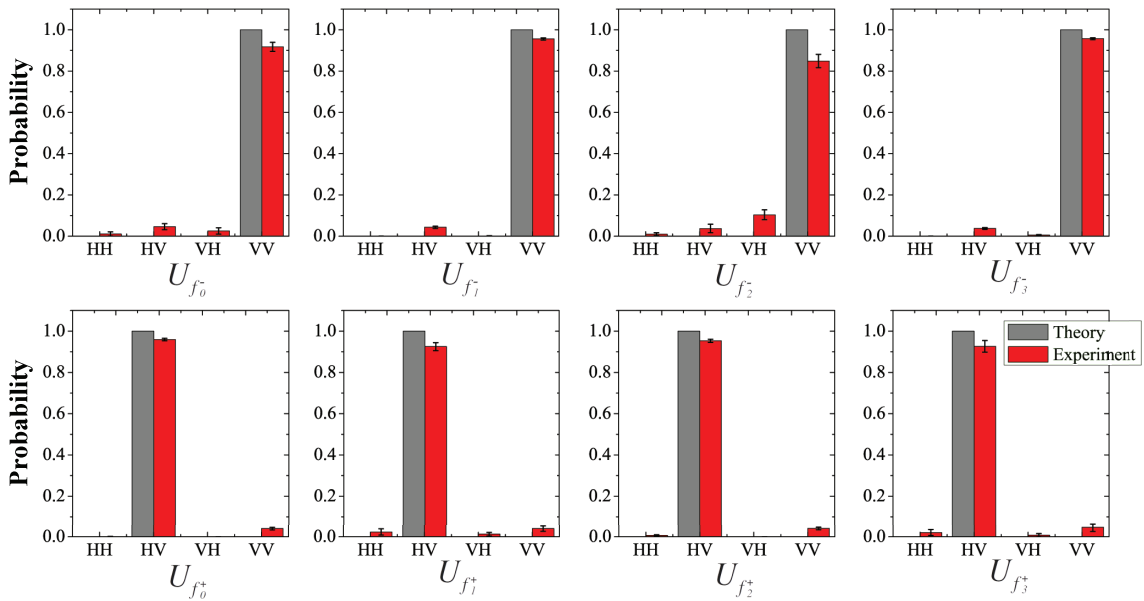}
 \caption{Experimental results: - Eight different permutations for four-dimensional set are chosen: $U_{f_{m}^{\pm}}$, where $m= 0, 1, 2, 3$. The quantum algorithm is to determine the parity of the permutations. $HH$, $HV$, $VH$, and $VV$ denote projection basis of the pair of photons. The outcome $HV$ means the parity of the permutation is positive, while the $VH$ tells the parity is negative (see main text for details). For eight possible permutations, the algorithm succeeds with the probability $93.023\pm2.015\%$ in average. The error bars denote the standard deviation calculated by the results of twenty times measurements.}
\label{fig:3}
\end{figure*}

Before running the algorithm, we characterize the property of the optical quantum circuit. Firstly, the two BDs form a Mach-Zehnder interferometer with a high visibility $99.691\pm0.004\%$, resulting from single photon counting. Secondly we prepare the photons input the first BD in $\ket{HV}$ state, and tune the angle of the HWP$_{2}$ to be $22.5^\circ$. In the two output mode, we observe a Hong-Ou-Mandel (HOM) interference of the two photons \cite{HOM}. With adjusting the optical path of the first photon, the visibility of HOM interference is $92.459\pm0.372\%$. At last, we prepare the state of pair of photons to be $(\ket{H}+\ket{V})\ket{V}/\sqrt{2}$ followed by a CNOT gate. In this case, the ideal output is a maximally entangled state $(\ket{HV}+\ket{VH})/\sqrt{2}$. Quantum state tomograph \cite{tomography} is used to determine the output of the CNOT gate, after which we obtain an entangled state with fidelity \cite{fidelity} $89.180\pm2.987\%$ compared to the ideal state.

We have realized the algorithm for all the eight possible permutations of four-dimensional set. The permutation operations are realized via tuning the HWP between the two BDs and two removable HWPs on the output modes. The outputs are injected into PBSs to be measured in the computational basis, and collected into single mode fibers. The photons are detected by avalanche photo-diodes (APDs) to give coincident counters.

We characterize the outputs by measuring the probabilities of each computational basis for each permutation. The ideal (gray bars) and measured (red bars) probabilities of all the permutations are shown in Fig. 3. The algorithm performed on the optical quantum circuit gives a probability of success $93.023\pm2.015\%$ in average.

{\it Conclusion and discussion}: - The original power of quantum computation is regarded as quantum correlations. However, our experiment proves that even a single pure qudit is sufficient to design an oracle-based algorithm which solves a black-box problem and demonstrates quantum speed-up over any classical approach to the same problem. We experimentally realized a quantum speed-up algorithm to determine the parity of permutation functions of four-dimensional set. This algorithm to determine the parity of a given permutation requires to call the permutation operation only once. Compared to evaluating the permutation function twice for classical algorithm, the quantum algorithm shows an intuitive speed-up. The optical quantum circuit performs quite well to give the successful probability $93.023\pm2.015\%$ in average. The experimental results are demonstrating the successful performance of the algorithm. The experiment has been performed in a photonic system, which, due to the strong potential of using photonics for advanced quantum information processing, makes our scheme ideal for probing of the boundary between classical and quantum efficiency in computing algorithms.

This algorithm for a single qudit to determine the parity of the permutation with only one evaluation of the function instead of two is so far the simplest quantum algorithm showing quantum speed-up. The initial motivation of realizing this algorithm is to show the role of contextuality in quantum speed-up. In our experiment, it appears that the speed-up mainly comes from the quantum parallelism, which is due to the supposition principle of quantum mechanics, together with another property of quantum mechanics called interference. Besides, the algorithm is deterministic and needs only one call, which makes it quite similar to Deutsch algorithm \cite{Nielsen,Deutsch}. Our experiment demonstrates the computing power of a single qudit by using only a simple toy algorithm. Deep analysis of this algorithm is not trivial, since it gives the possibility of understanding the relation between quantum supposition and contextulity, and this may also help to understand the relation between supposition and other candidates for the origin of the power of quantum computing, such as quantum correlations.

XZ thank Jin-Shi Xu for helpful guidance in experiment. HQ thank Kai Sun for useful discussion about analysis of data. This work has been supported by the National Natural Science Foundation of China under Grant Nos. 11174052 and 11474049, the Open Fund from the State Key Laboratory of Precision Spectroscopy of East China Normal University, the National Basic Research Development Program of China (973 Program) under Grant No. 2011CB921203 and CAST Innovation fund.

%
\end{document}